\documentclass[12pt]{article}
\usepackage{amsmath,amssymb,amsbsy,amsfonts,amsthm,latexsym,amsopn,amstext,
            amsxtra,euscript,amscd}
\title{{\large\bf Conditions on the generator for forging ElGamal signature}}
\author{Omar Khadir\\
 Laboratory of Mathematics, Cryptography and Mechanics,\\Department of  Mathematics, Fstm\\
  University of Hassan II-Mohammedia, Morocco\\
       }

\date{}
\begin{document}
\maketitle

 \baselineskip=18pt

\noindent\hrulefill

{\abstract \small This paper describes new conditions on
parameters selection that lead to an efficient algorithm for
forging ElGamal digital signature. Our work is inspired by
Bleichenbacher's ideas.}

\noindent\hrulefill

 \vspace{0.3cm} \noindent {\small \bf Keywords :} \
{\small Public key cryptography, discrete logarithm problem,
ElGamal digital signature.}

\vspace{0.5cm}

\noindent {{\small \bf MSC 2010 : } {\small 94A60}

 \section{Introduction}
Numerous digital signature algorithms have been developed since
the invention of the public key cryptography in the late
1970s[3,15,14]. They almost all have the same principle. Every
user possesses two kinds of keys. The first one is private, must
be kept secret and stored only locally. The second is public and
must be largely diffused to be accessible to the others users. To
sign a particular message, a contract or a will $M$, Alice has to
solve a hard mathematical equation depending of $M$ and of her
public key. With the help of her private key, she is able to
furnish the solutions. Bob, the judge or anybody, can verify that
the solutions computed  by Alice are valid. For an adversary,
without knowing Alice private key, the algorithm is constructed in
such a way that it is computationally too hard to solve the
considered equation.\\
One of the most popular signature algorithm was proposed by
ElGamal[4]. It has many variants[16,17,9,5] and is based on the
hard discrete logarithm problem. Since its conception in 1985,
several attacks were mounted and have revealed possible weaknesses
if the signature keys were not carefully selected[2,1,10,13].
However, no general method
for breaking totally the system is known.\\
In an ElGamal signature protocol, a signer, in addition to his
private key, must detain three other integer parameters
$(p,\alpha,y)$ as a public key. In 1996, Bleichenbacher[2]
presented a cryptanalysis where he showed that if the generator
$\alpha$ and the modulus $p$ verify  some special relations, it is
possible to forge ElGamal signature for any arbitrary message. In
particular, he proved that, the signature scheme becomes insecure
when parameters $\alpha$ and $p$
are chosen such that $\alpha$ divides $p-1$. Hence, selecting  $\alpha=2$ is imprudent.\\
 The purpose of our work, is to describe new conditions on
parameters selection that lead to an efficient algorithm for
forging ElGamal signature for any arbitrary message. As an
extension of Bleichenbacher's result, we show that, if the modular
inverse of the generator $\alpha$ divides $p-1$, then it is
possible to break the system. As an example, the choice of
$\displaystyle \alpha=\frac{p+1}{2}$ as a generator, is not
recommended.
\\
The paper is organized as follows. Section 2 contains
preliminaries which will be utilized in the sequel. Our
contribution, mainly composed by
Algorithm 2 and Corollary 3, is presented in section 3. We conclude in section~4. \\
Throughout this article, we will adopt  ElGamal paper
notations[4]. $\mathbb{Z}$, $\mathbb{N}$ are respectively the sets
of integers and non-negative integers. For every positive integer
$n$, we denote by $\mathbb{Z}_n$ the finite ring of modular
integers and by $\mathbb{Z}_n^*$ the multiplicative group of its
invertible elements. Let $a,b,c$  be three integers. The great
common divisor of $a$ and $b$ is denoted by $gcd(a,b)$.  We write
$a\equiv b$ $[c]$ \ if $c$ divides the difference $a-b$, and $a=b\
mod\ c$ if $a$ is the remainder in the division of $b$ by $c$. The
positive integer $a$ is said to be B-smooth[8, p.92],
$B\in\mathbb{N}$,  if
every prime factor of $a$ is less than or equal to the bound $B$.\\
 We start, in the next section, by preliminaries containing known mathematical facts
 that will be exploited later.
\section{Preliminaries}
Before exploring new situations under which one can forge ElGamal
digital signature, we briefly review three questions that are
directly related to our result.
\subsection{Discrete logarithm problem when $p-1$ is B-smooth}
The discrete logarithm problem importance started to grow with the
publication in 1976 of the fundamental work of Diffie and
Hellman[3,11]. The issue became central in public key
cryptography.\\ Let $p$ be a prime integer and $\alpha$ a
primitive root of $\mathbb{Z}_n^*$. We consider the discrete
logarithm equation
\begin{equation}
 \alpha^x\equiv y\ [p]
\end{equation}
 where $y$ is fixed  in $\{1,2,3,\ldots,p-1\}$, and $x$ is  unknown in $ \{0,1,2,\ldots,p-2\}$.\\
 In 1978, Pohlig and Hellman[12] published a practical  method to solve equation (1) when all the prime factors of $p-1$ are not too
 large. Let us recall the outlines of their algorithm.\\
 Assume that $p-1$ is B-smooth. The bound B depends on the computers power. This implies that we can obtain the prime factorization of $p-1$ :
 $p-1=p_1^{n_1}\,p_2^{n_2}\, \ldots \, p_k^{n_k}$
 where  $n_i,k_i\in \mathbb N^*$ for $1\leq i\leq k$. We will first find $x$ modulo $p_i^{n_i}$ for every $i\in\{1,2,\ldots, k\}$
 and apply the Chinese Remainder Theorem[8, p.68] to compute $x$ modulo $p_1^{n_1}\,p_2^{n_2}\, \ldots \,
 p_k^{n_k}$.\\
The $p_1$-ary representation  of \ $x_1=x\ mod\ p_1^{n_1}$ \ can
be written as :\begin{equation} x_1=b_0+b_1\, p_1+\,
\ldots+b_{n_1-1}\, p_1^{n_1-1}
\end{equation}
where $b_0,b_1,\ldots,b_{n_1-1}$ are unknown in $\{0,1,\ldots p_1-1\}$.\\
 Let $\lambda_1=p_1^{n_1-1}\,p_2^{n_2}\, \ldots \, p_k^{n_k}$. We
 have $\lambda_1\,x_1=\lambda_1\,b_0+K_1\,(p-1),\ K_1\in\mathbb N$.
Since $\alpha^{K_1\,(p-1)}\equiv 1\ [p]$,  equation (1) can be
transformed to $\alpha^{\lambda_1\,x_1}\equiv y^{\lambda_1\,}\
[p]$ and therefore
\begin{equation}
\alpha^{\lambda_1\,b_0}\equiv y^{\lambda_1}\ [p]
\end{equation}
From equation (3) we obtain the first coefficient $b_0$.\\
Similarly, if $\lambda_2=p_1^{n_1-2}\,p_2^{n_2}\, \ldots \,
p_k^{n_k}$, then
$\lambda_2\,x_1=\lambda_2\,b_0+\lambda_2\,b_1\,p_1+K_2\,(p-1),\
K_2\in\mathbb N$. As  $\alpha^{K_2\,(p-1)}\equiv 1\ [p]$, equation
(1) becomes $\alpha^{\lambda_2\,x_1}\equiv y^{\lambda_2}\ [p]$ and
therefore
\begin{equation}
\alpha^{\lambda_2\,b_0+\lambda_2\,b_1\,p_1}\equiv y^{\lambda_2}\
[p]
\end{equation}
We get the second coefficient $b_1$ from equation (4).\\
Gradually, we compute $b_0,b_1,b_2,\ldots, b_{n_1-1}$ and then
determine $x_1$. This is computationally possible since
$2^{n_1}\leq p-1$ and then $\displaystyle n_1\leq \frac{\ln
(p-1)}{\ln 2}$.
In other word, $n_1$ is bounded by the bit length of $p-1$. \\
We repeat the technique with $p_2,p_3,\ldots,p_k$ and  arrive to
the following
system  of congruences :\\
\begin{equation}
\left\{%
\begin{array}{cc}
  x\equiv x_1 & [p_1^{n_1}] \\
  x\equiv x_2 & [p_2^{n_2}] \\
  \vdots & \vdots \\
  x\equiv x_k & [p_k^{n_k}] \\
\end{array}%
\right. \end{equation} Natural index $k$ is not too large since
$\displaystyle k\leq \sum_{i=1}^k\, n_i\leq \frac{\ln (p-1)}{\ln
2}$. Hence, system (5) can efficiently be solved by the Chinese
Remainder Theorem method whose running time is $O(\ln^2 p)$ bit
operations. The complexity of Pohlig-Hellman algorithm is
$O(\sum_{i=1}^k \, n_i(\ln p +\sqrt{p_i}))$ bit operations[12, 9,
p.108].
\subsection{ElGamal signature algorithm}
We recall the basic ElGamal protocol[4,19,8] in three steps.

\vspace{0.2cm} \noindent {\bf 1.} Alice begins by choosing three
numbers $p,\alpha$ and $x$ such that :

 \hspace{1cm} - $p$ is a large prime integer.

\hspace{1cm} - $\alpha$ is a primitive root of the finite
multiplicative group $\mathbb Z_p^*$.

\hspace{1cm} - $x$ is a random element taken in
$\{1,2,\ldots,p-2\}$.

\noindent She computes  $y=\alpha^x\ mod\ p$ and publishes the
triplet $(p,\alpha,y)$ as her public key. She keeps secret the
parameter $x$ as her private key.

\vspace{0.2cm} \noindent
 {\bf 2.} Suppose that Alice desires to  sign the message  $m<p$. She must
 solve the modular equation
\begin{equation}
 \alpha^m \equiv y^r\,r^s\ [p]
\end{equation}
where $r$ and $s$ are two unknown variables. \\
Alice computes $r=\alpha^k\ mod\ p$, where $k$ is selected
randomly and is invertible modulo $p-1$. She has exactly
$\varphi(p-1)$ possibilities for $k$, where $\varphi$ is the
phi-Euler function. Equation (6) is then equivalent to :
\begin{equation}
 m\equiv x \, r+k\, s \ [p-1]
\end{equation}
As Alice possesses the secret key $x$, and as the integer $k$ is
invertible modulo $p-1$, she  computes the second unknown variable
$s$ from relation (7) by :
\begin{equation}
\displaystyle s\equiv \frac{m-x\, r}{k}\ [p-1]
\end{equation}
The inverse modulo $p-1$ of the integer $k$ in equation (8) is
computed by the  extended Euclidean algorithm whose complexity is
$O(\ln^2 p)$ bit operations.

 \vspace{0.2cm} \noindent
 {\bf 3.} Bob can verify the signature by checking  that congruence
 (6) is valid.

\vspace{0.4cm} \noindent  Observe that, in step {\bf 1.}, we need
to know how to construct signature keys. Generally, the running
time for generating prime integers takes the most important part
in the total running time. In [6], we made experimental tests and
concluded by suggesting some rapid procedures. In step {\bf 2.},
the random integer $k$ must be kept secret, otherwise relation (7)
allows any adversary to obtain Alice private key $x$. The fact of
having many possibilities for the valid pairs $(r,s)$ does not
affect the system security. Indeed, these pairs are uniformly distributed[18]. \\
To prevent obvious attacks  against ElGamal signature scheme, some
of theme mentioned in the original paper[4], it is necessary to
work with a free collision hash function $h$. The message $M$ is
simply replaced by $m=h(M)$ before applying the signature
algorithm. We can take $h$ equal to the secure hash algorithm
SHA1[19. p139, 8. chap9].
\subsection{Bleichenbacher's attack}
In Eurocrypt'96 meeting, Bleichenbacher indicated  a possible
weakness in the ElGamal signature scheme if the keys are not
properly chosen[2,1]. His ideas are summarized in his main
theorem.

\vspace{0.4cm}\noindent
 {\bf Theorem 1.[2]} \ Let $p-1=b\,w$ where $b$ is
B-smooth and let $y_A\equiv \alpha^{x_A}\ [p]$ be the public key
of user $A$. If a generator $\beta=c\, w$ with $0<c<b$ and an
integer $t$ are known such that $ \beta^t\equiv \alpha \ [p]$ then
a valid signature $(r,s)$ on a given $h$ can be found.

 \vspace{0.4cm}\noindent
\noindent This theorem has the immediate consequence :

 \vspace{0.4cm}\noindent {\bf
Corollary 1.[2]} \ If $\alpha$ is B-smooth and divides $p-1$, then
it is possible to generate a valid ElGamal signature on an
arbitrary value $h$.

\vspace{0.3cm} \noindent
 Now, we can move to the next section where we expose our contribution.
 \section{Our contribution}
In this section, we present our main result. The first sufficient
condition for forging ElGamal signature is based on a slight
simplification of Bleichenbacher's theorem. More precisely, we
have :

\vspace{0.3cm} \noindent
 {\bf Theorem 2.} Let $(p,\alpha,y)$ be Alice public key in an ElGamal signature scheme.
 If an adversary can compute
 a nonnegative integer $k\leq p-2$, relatively prime to
$p-1$ and such that $\displaystyle \frac{p-1}{gcd(p-1,\alpha^k\
mod \ p)}$ is B-smooth, then he will be able to forge Alice
signature. \proof We follow the method used in [2]. Let
$(p,\alpha,y)$ be Alice public key in an ElGamal signature
protocol. If we put $D=gcd(p-1,\alpha^k\ mod\ p)$, then there
exist two coprimes $p_1$ and $a_1$ such that $p-1=D\, p_1$ and
$\alpha^k\ mod\ p=D\, a_1$. Let $H$ be the subgroup of
$\mathbb{Z}_p^*$ generated by the particular element $\alpha^D\,
mod \ p$. Since $y\equiv \alpha^x\, [p]$, where the natural
integer $x$ is Alice secret key, we have $y^D\equiv (\alpha^D)^x\,
[p]$ and then $y^D\, mod\, p\in  H$. By hypothesis, the order of
the subgroup $H$, $\displaystyle \frac{p-1}{D}=p_1$ is B-smooth,
so the discrete logarithm equation
\begin{equation}
(\alpha^D)^X\equiv y^D\, [p]
\end{equation}
can be solved in
 polynomial time. Hence there exists $x_0\in \mathbb N$ such that
$(\alpha^D)^{x_0}\equiv y^D\, [p]$. To forge Alice signature for a
message $M$, and a hash function $h$, if the adversary puts
$m=h(M)$, he must find two positive integers $r,s$ such that
$\alpha^m\equiv y^r\, r^s\, [p]$. If he chooses $r=\alpha^k\,
mod\, p$, he will have the
equivalences~:\\
 $\alpha^m\equiv y^r\, r^s\,[p]\Longleftrightarrow$ $
\alpha^m\equiv y^{D\,a_1}\,\alpha^{k\,s}\, [p]\Longleftrightarrow$
$\alpha^m\equiv [(\alpha^D)^{x_0}]^{a_1}\, \alpha^{k\,s}\,
[p]\Longleftrightarrow$ $ \alpha^m\equiv \alpha^{x_0\,
r}\,\alpha^{k\,s}\,
[p]$\\
$\Longleftrightarrow m\equiv x_0\, r+k\,s\, [p-1]$ $\displaystyle
\Longleftrightarrow s\equiv
\frac{m-x_0\,r}{k}\, [p-1].$\\
So $(r,s)$ is a valid signature for the message $M$, obtained
without knowing Alice secret key.

\hspace{14.5cm} $\qed$

\vspace{0.4cm} \noindent If the first valid exponent $k$ in our
Theorem 2 is not too large, the adversary can  construct the
following deterministic algorithm in order to forge ElGamal
signature. Consequently, we recommend that when selecting the
signature keys, we have to verify that for any $k\leq K_0$, where
$gcd(k,p-1)=1$ and  $K_0$ is the largest bound allowed by computer
power, the integer $\displaystyle \frac{p-1}{gcd(p-1,\alpha^k\ mod
\ p)}$ has at least one large prime factor.

\vspace{0.7cm}\noindent {\bf Algorithm 1}\\
{\bf Input :} Alice public key $(p,\alpha,y)$ and the message $M$ to be signed.\\
{\bf Output :} The signature $(r,s)$ of $M$.

\vspace{0.3cm}\noindent
{\bf 1.} Read($p,\alpha,y)$; \{$(p,\alpha,y)$ is Alice public key\}. \\
{\bf 2.} Read($M$); m:=h(M); \{$m$ is  the hashed of the message to be signed\}. \\
{\bf 3.} $j\leftarrow -1$; \{Initialization of integers $j$ which play the role of exponents $k$\}.\\
 {\bf 4.} $F\leftarrow 0$; \{$F$ is a flag\}.\\
{\bf 5.} While (F=0) do

\hspace{1cm} {\bf 5.1.} $j\leftarrow j+2;$ \{$j$ must be
invertible modulo $p-1$, so $j$ is odd\}.

\hspace{1cm} {\bf 5.2.} If $gcd(p-1,j)=1$ then

\hspace{3cm} {\bf 5.2.1.} $r\leftarrow \alpha^j\  mod\ p;$ \{ $r$
will be the first parameter of the signature\}.

\hspace{3cm} {\bf 5.2.2.} $D\leftarrow gcd(p-1,r)$; \{ $D$ will be
the exponent in equation (9)\}.

\hspace{3cm} {\bf 5.2.3.} If $(p-1)/D$ is B-smooth, then

\hspace{5cm} {\bf 5.2.3.1} $k\leftarrow j$; \{ $k$ is the searched
exponent of $\alpha$\}.

\hspace{5cm} {\bf 5.2.3.2} $F\leftarrow 1$; \{ To stop the while loop\}.\\
 {\bf 6.} $x_0\leftarrow X$; \{$X$ is a solution of equation
(9), obtained by Pohlig-Hellman algorithm[12]\}.\\
{\bf 7.}  $\displaystyle s\leftarrow \frac{m-x_0\, r}{k}\ mod\
(p-1)$;
\{$s$ is the second parameter of the signature\}.\\
{\bf 8.} Return$(r,s)$; \{$(r,s)$ is the digital signature\}.

 \vspace{0.4cm}\noindent

\vspace{0.5cm}\noindent Our theorem 2 has a first remarkable
consequence : if $Q$ denotes the part of the prime factorization
of $p-1$ that is not B-smooth and if $\alpha^k \ mod \ p$, for
some $k\in \mathbb N$, is a multiple of $Q$ then ElGamal signature
scheme is insecure. More formally :

\vspace{0.4cm}\noindent {\bf Corollary 2.} Let
$p_1^{n_1}\,p_2^{n_2}\ldots p_k^{n_k}\,
q_1^{n'_1}\,q_2^{n'_2}\ldots q_l^{n'_l}$ be the classical prime
factorization of $p-1$, where
$p_1^{\alpha_1}\,p_2^{\alpha_2}\ldots p_r^{\alpha_r}$ is B-smooth.
If an adversary  can compute a natural integer $k$, $k\leq p-2$,
relatively prime to $p-1$ and such that $\alpha^k\ mod \ p$ is a
multiple of $Q=q_1^{n'_1}\,q_2^{n'_2}\ldots q_l^{n'_l}$, then he
will be able to forge Alice signature for any arbitrary message
$M$. \proof Observe, first, that $\alpha$ is not necessary a
divisor of $p-1$, nor a B-smooth integer. Since $\alpha^k\ mod \
p$ is a multiple of $Q$, we have $Q$ divides $gcd(p-1,\alpha^k\
mod \ p)$. So $p_1^{n_1}\,p_2^{n_2}\ldots p_k^{n_k}$ is a multiple
of $\displaystyle \frac {p-1}{gcd(p-1,\alpha^k\ mod \ p)}$,  and
then $\displaystyle \frac {p-1}{gcd(p-1,\alpha^k\ mod \ p)}$ is
B-smooth. We

 \vspace{0.3cm}\noindent
conclude by applying Theorem 2.

 \hspace{14.5cm} $\qed$

\noindent If the first valid exponent $k$ is not too large,
Corollary 2 leads to a more practical algorithm for forging
ElGamal signature.

\vspace{0.4cm}\noindent {\bf Algorithm 2}\\
{\bf Input :} Alice public key $(p,\alpha,y)$ and the message $M$ to be signed.\\
{\bf Output :} The signature $(r,s)$ of $M$.

\vspace{0.3cm}\noindent
{\bf 1.} Read($p,\alpha,y)$; \{$(p,\alpha,y)$ is Alice public key\}. \\
{\bf 2.} Read($M$); m:=h(M); \{$m$ is  the hashed of the message to be signed\}. \\
{\bf 3.} $Q_0\leftarrow Q$; \{$Q$ is the part of $p-1$ that is not B-smoot\}. \\
{\bf 4.} $j\leftarrow -1$; \{Initialization of integers $j$ which play the role of exponents $k$\}.\\
 {\bf 5.} $F\leftarrow 0$; \{$F$ is a flag\}.\\
{\bf 6.} While (F=0) do

\hspace{1cm} {\bf 6.1.} $j\leftarrow j+2;$ \{$j$ must be
invertible modulo $p-1$, so $j$ is odd\}.

\hspace{1cm} {\bf 6.2.} If $gcd(p-1,j)=1;$ then

\hspace{3cm} {\bf 6.2.1.} $r\leftarrow \alpha^j\  mod\ p;$ \{ $r$
will be the first parameter of the signature\}.

\hspace{3cm} {\bf 6.2.2.} If $r\ mod\  Q_0=0$ then

\hspace{5cm} {\bf 6.2.2.1.} $k\leftarrow j$; \{ $k$ is the
searched exponent of $\alpha$\}.

\hspace{5cm} {\bf 6.2.2.2.} $F\leftarrow 1$; \{ To stop the while loop\}.\\
 {\bf 7.}  $D\leftarrow gcd(p-1,r)$; \{ $D$ will be the exponent in equation (9)\}. \\
{\bf 8.} $x_0\leftarrow X$; \{$X$ is a solution of equation (9), obtained by Pohlig-Hellman algorithm[12]\}.\\
{\bf 9.} $\displaystyle s\leftarrow \frac{m-x_0\, r}{k}\ mod\
(p-1)$;
\{$s$ is the second parameter of the signature\}.\\
{\bf 10.} Return$(r,s)$; \{$(r,s)$ is the digital signature\}.

 \vspace{0.4cm}\noindent
Next result, which can be seen as an extension of Blechenbacher's
Corollary 1, shows that in an ElGamal signature scheme, it is not
secure to have a primitive root whose modular inverse divides
$p-1$. In particular, as a primitive root, $\displaystyle
\alpha=\frac{p+1}{2}$ is not recommended  since its inverse is 2.
More explicitly :

 \vspace{0.3cm}\noindent {\bf
Corollary 3.} Let $(p,\alpha,y)$ be Alice public key in an ElGamal
signature protocol. An adversary can forge Alice signature for any
given message if one of the following conditions is satisfied :\\
a) $p\equiv 1\ [4]$, $\alpha$ is B-smooth and divides $p-1$. \\
b) $p\equiv 1\ [4]$, $\displaystyle \frac{1}{\alpha}\ mod\ p$ is B-smooth and divides $p-1$. \\
c) $\alpha^2$ is B-smooth and divides $p-1$. \proof a) Put
$p-1=\alpha\,  Q$. As $\alpha$ is a primitive root, we have
$\alpha^{(p-1)/2}\equiv -1\ [p]$, and so $\alpha^k\equiv Q\ [p]$
where $k=(p-3)/2$. Consequently $gcd(p-1,\alpha^k\  mod\ p)=Q$ and
then $\displaystyle \frac{p-1}{gcd(p-1,\alpha^k\  mod\
p)}=\alpha$ which is smooth and this allows the use of our theorem 2. \\
b) It is easy to see that $\alpha$ is a primitive root modulo $p$
if and only if $\displaystyle \frac{1}{\alpha}\ mod\ p$ is a
primitive root. Suppose that $\displaystyle \frac{1}{\alpha}\ mod\
p$ is B-smooth and divides $p-1$. If the public key was
$(p,\displaystyle \frac{1}{\alpha}\ mod\ p,\displaystyle
\frac{1}{y}\ mod\ p)$, and the private key was the same parameter
$x$, an adversary would be able to forge the signature and to find
two valid integers $(r_1,s_1)$ for any arbitrary message $M$. With
$(r,s)=(r_1,-s_1\ mod\ p-1)$, the adversary forges Alice signature
for the message $M$. Indeed, ElGamal equation (1) is equivalent to
\begin{equation}
\displaystyle (\frac{1}{\alpha})^m\equiv (\frac{1}{y})^{r_1}\,
(r_1)^{-s_1}\ [p]
\end{equation}
c) If $p\equiv 1\ [4]$, then the affirmation is true from case a).
Assume then that $p\equiv 3\ [4]$ and put $p=3+4K$, $K\in \mathbb
N^*$. As $\alpha$ is a primitive root, we have
$\alpha^{(p-1)/2}\equiv -1\ [p]$, and then $\alpha^2\,
(\alpha^{(p-5)/2}\  mod\  p)=p-1$. Let $k=(p-5)/2$. Since
$k=2K-1$, $gcd(p-1,k)$ divides 4 and as $k$ is odd,
$gcd(p-1,k)=1$. On the other hand $\displaystyle \frac
{p-1}{gcd(p-1,\alpha^k\ mod \ p)}=\frac {p-1}{\alpha^k\ mod \
p}=\alpha^2$ is B-smooth which allows us to apply theorem 2 \ and
achieve the proof.

\hspace{14.5cm} $\qed$

 \vspace{0.5cm}\noindent
Before concluding, we give the following theoretical theorem
relative to the number of exponents $k$ figuring in Corollary 2
and Algorithm 2.

  \vspace{0.5cm}\noindent {\bf
Theorem 3.} Let $\alpha$ be a primitive root of the multiplicative
group $\mathbb Z_p^*$. For any fixed integer $Q$ such that $1\leq
Q\leq p-1$, if we set
\begin{equation}
E_\alpha=\{k\in \mathbb N\ /\ 1\leq k\leq p-2,\  gcd(p-1,k)=1,\
{\rm and}\ Q\ {\rm divides}\  \alpha^k\ mod\ p\}
\end{equation}then the cardinality of $E_\alpha$ is independent of
the choice of the primitive root $\alpha$.

\proof Let $\alpha,\ \beta$ be two fixed primitive roots of
$\mathbb Z_p^*$. It is well-known that there exists
$i\in\{1,2,\ldots,p-2\}$ such that $gcd(p-1,i)=1$  and
$\alpha\equiv \beta^i \ [p]$. Consider then the function~:

\hspace{4cm} $f\  :\ E_\alpha\longrightarrow E_\beta$

\hspace{5cm} $k \mapsto ik\ mod \ (p-1)$

\noindent First we have $f(k)\in E_\beta$, where $E_\beta$ is
defined like $E_\alpha$ in relation (13). Indeed $gcd(p-1,ik\
mod\ (p-1))=gcd(p-1,ik)=1$ since $gcd(p-1,k)=gcd(p-1,i)=1$. \\
On the other hand $\beta^{ik\ mod\ (p-1)}\equiv \beta^{ik}\equiv
\alpha^k\ [p]$, so $Q$ divides $\beta^{ik\ mod\ (p-1)}\ mod\ p$
and therefore $ik \ mod \ (p-1)\in
E_\beta$.\\
Let us now establish that $f$ is an injective function. We have successively :\\
$f(k)=f(k')\Longrightarrow ik\ mod \ (p-1)=ik'\ mod \
(p-1)\Longrightarrow i\, k \equiv ik'\ [p-1]\Longrightarrow
k\equiv k'\ [p-1]$ since $i$ is invertible modulo $p-1$. As $1\leq
k,k'\leq p-2$, we obtain that
$k=k'$.\\
Since $f$ is injective $Card(E_\alpha)\leq Card(E_\beta)$ and by
interchanging the role of the parameters  $\alpha$ and $\beta$, we
find that $Card(E_\beta)\leq Card(E_\alpha)$ and so
$Card(E_\alpha)= Card(E_\beta)$. This means that $Card(E_\alpha)$
is a constant depending only of the two integers $p$ and $Q$.

\hspace{14.5cm} $\qed$
\section{Conclusion}
In this paper, we described new conditions on parameters selection
that can lead to an efficient deterministic algorithm for forging
ElGamal digital signature. Our approach is based on the work of
Bleichenbacher
presented at Eurocrypt'96 conference[2].\\

\end{document}